\documentclass[a4paper,fleqn,usenatbib]{mnras}

\usepackage{mathptmx}

\usepackage[T1]{fontenc}
\usepackage{ae,aecompl}

\usepackage{graphicx}	
\usepackage{amsmath}	
\usepackage{amssymb}	
\usepackage{rotating}

\newcommand{\msun}{M_\odot}

\newcommand{\mc}{M^*}

\newcommand{\ms}{M_\star}

\newcommand{\Mp}{M_{\rm p}}
\newcommand{\Rp}{R_{\rm p}}
\newcommand{\Reff}{R_{\rm e}}
\newcommand{\sige}{\sigma_{\rm e}}
\newcommand{\SBe}{\mu_{\rm e}}
\newcommand{\ti}{t_{\rm i}}

\newcommand{\zi}{z_{\rm{i}}}

\newcommand{\zo}{z_{\rm{obs}}}
\newcommand{\Om}{\Omega_{\rm m}}

\newcommand{\deli}{\delta_{\rm i}}
\newcommand{\rhoc}{\rho_{\rm{crit}}}

\newcommand{\Ngal}{N_{\rm gal}}
\newcommand{\Rd}{R_{\rm d}}

\title[The formation of first-ranked galaxies]{Unveiling the formation
  route of the largest galaxies in the universe}

\author[J.~D.\ Perea and J.~M.\ Solanes]{\parbox{\textwidth}{
Jaime D.\ Perea$^{1}$\thanks{E-mail: \texttt{jaime@iaa.es}}
and Jos\'e M.\ Solanes$^{2}$}\vspace{0.4cm}
\\
$^{1}$Instituto de Astrof\'\i sica de Andaluc\'\i a, IAA--CSIC. Glorieta de la Astronom\'\i a, s/n; E--18008~Granada, Spain\\
$^{2}$Departament de F\'\i sica Qu\`antica i Astrof\'\i sica and Institut de Ci\`encies del Cosmos (ICCUB), Universitat de Barcelona.\\ 
\ C.\ Mart\'{\i} i Franqu\`es, 1; E--08028~Barcelona, Spain
}

\date{Accepted to MNRAS} 

\pubyear{2016}

\begin{document}
\label{firstpage}
\pagerange{\pageref{firstpage}--\pageref{lastpage}}
\maketitle

\begin{abstract}
Observational evidence indicates that the role of gas is secondary to
that of gravity in the formation of the most luminous spheroids
inhabiting the centres of galaxy associations, as originally
conjectured in the late 80's/early 90's. However, attempts to explain
the origin of the Fundamental Plane (FP) of massive early-type
galaxies (ETGs) -- a tilted version of the scaling relation connecting
the size, velocity dispersion and mass of virialized homologous
systems -- based on sequences of pairwise mergers, have systematically
concluded that dissipation cannot be ignored. We use controlled
simulations of the previrialization stage of galaxy groups to show
that multiple collisionless merging is capable of creating realistic
first-ranked galaxies. Our mock remnants define a thin FP that
perfectly fits data from all kinds of giant ETGs in the local volume,
showing the existence of a unified relationship for these
systems. High-ranked galaxies occupy in the FP different areas than
standard objects, a segregation which is viewed essentially as
zero-point offsets in the 2D correlations arising from standard
projections of this plane. Our findings make a strong case for
considering hierarchical dissipationless merging a viable route for
the formation of the largest galaxies in the universe.
\end{abstract}

\begin{keywords}
galaxies: elliptical and lenticular, cD -- galaxies: formation --
galaxies: fundamental parameters -- galaxies: groups: general --
galaxies: interactions -- galaxies: kinematics and dynamics
\end{keywords}



\section{The mergers-make-ellipticals hypothesis}\label{intro}

The possibility that ellipticals form by mergers of two or more
pre-existing galaxies, typically discs, was first suggested by
\citet{Too77} as an alternative to the top-down monolithic collapse
hypothesis \citep*{EL-BS62}. That was some years before cosmological
observations start to point heavily to a world model in which galaxies
form within an underlying structure built up hierarchically through
the collapse and subsequent merger of Cold Dark Matter (CDM)
haloes. It was precisely this generic feature of CDM models, the
continuous bottom-up growth of structure in all scales -- nearly
unabated until recently ($z\lesssim 0.5$) when the universe has
entered a phase of dark energy domination --, that provided important
clues to alleviate the first objections raised by early simulations of
the merger scenario. In those days, numerical models mostly focused on
pairwise collisions of similar stellar discs, producing remnants that
rotated too fast and had ellipticities that were too large to
represent true massive ellipticals. The addition of an extended CDM
component around the colliding galaxies facilitated the transfer of
the orbital angular momentum deposited in the discs along the merger
to the common dark halo, thus reducing the rotation speed of the final
stellar remnant to values consistent with observations. Moreover, it
soon became clear that understanding the formation of early-type
galaxies (ETGs) demanded repeated mergers of progenitors covering a
wide range of masses
\citep*[e.g.][]{She03,CLV07,TFdR11,Moo14}. Multiple merging led to the
formation of oblate-triaxial, and therefore more rounder-looking,
remnants \citep{WH96}.

Another important obstacle to the merger hypothesis is the necessity
of building objects exhibiting central light distributions more
concentrated that those of their most likely progenitors. As stated by
\citet{Car86}, the macroscopic phase space densities of the cores of
intermediate- and low-luminosity elliptical galaxies are found to
generally exceed the phase space densities measured in galactic
discs. Since according to Liouville's theorem the phase space
probability in a dynamically reversible system must be kept constant,
this argues against the formation of the bulk of the elliptical
population via collisionless mergers of stellar discs, a point also
stressed by \citet{Ost80} soon after the merger hypothesis was
formulated. To overcome this problem \citeauthor{Car86} suggested
dealing with mergers of spirals that either include a high-mass
density spheroidal component (e.g.\ a bulge), or are largely gaseous
and thus capable of introducing irreversibility in the system by
effectively dissipating part of the energy deposited into the
core. The fact that the central space densities of ellipticals depend
strongly on luminosity \citetext{$\propto L_B^{-2.35}$ as estimated by
\citealt{Car86}} turned dissipative merging into an alternative that
was worth exploring.

Using a large suite of simulations of equal-mass mergers of discs,
both with and without gas, \citet{Rob06} \citep*[see also][]{HCH08},
showed that dissipation imprints a mass-dependent relation in the
ratio of total mass to stellar mass of the remnants of the form
\begin{equation}\label{MMstar}
\frac{M}{\ms}\propto \ms^\gamma\;.
\end{equation}
This relationship arises from the halo mass dependence of the cooling
efficiency of the intergalactic medium (IGM) collisionally heated
during the merger. In gas-rich low-mass systems the hot intergalactic
gas cools rapidly producing inflows giving rise to a central merger
induced starburst that is compact. However, in a wet merger between
two large spirals the IGM cannot cool efficiently and remains in the
halo for the duration of the merger, causing more massive galaxies to
act increasingly as dissipationless systems. This produces low- and
intermediate-mass final remnants that are more baryon dominated in
their central regions than their more massive counterparts, giving
rise to a dependence of the form~(\ref{MMstar}) with $\gamma$ small
($\sim 0.1$--$0.2$) but positive\footnote{The evidence supporting a
  positive slope for the elliptical population is still controversial,
  especially from the data recently reported by the ATLAS3D survey
  \citep[][]{Cap13}.}, as found in some observational studies
\citep[e.g.][]{Pad04}.

It is straightforward to show, provided $R\propto \ms^\mu$, that
equation~(\ref{MMstar}) can be transformed into the power-law relation
\begin{equation}\label{FPt}
R\propto (\sigma^2I^{-1})^\lambda\;,
\end{equation}
with $I\propto \ms/R^2$, $R$ the projected radius of the system,
$\sigma$ its line-of-sight velocity dispersion\footnote{In the present
  reasoning we are assuming, for simplicity, sphericity and isotropic
  orbital structure for the galaxies. For a more general discussion of
  the FP and its relation with the virial theorem see, for instance,
  \citet*{Cio97} and references therein.} and
\begin{equation}\label{lambda}
\lambda=\left(\frac{\gamma}{\mu}+1\right)^{-1}\;,
\end{equation}
an index measuring the ``tilt'' relative to the standard plane,
$R\propto\sigma^2I^{-1}$, that arises when one assumes for virialized
remnants baryon fraction invariance and structural and dynamical
homology. The most useful version of equation~(\ref{FPt}) is none the
less a generalization that accounts for either nonhomology, or
variations in the the stellar mass fraction, or both, and that
replaces $\ms/R^2$ by a directly measurable property such as the
surface brightness $\mu$. If we transform this new expression by
dividing it by $R^2$ and taking logarithms, solving for $R$ we get
\begin{equation}\label{logFP}
\log R = a\log\sigma+b(\mu/2.5) + c\;.
\end{equation}
In this case the tilt is expressed in slope coefficients that differ
from the standard values: $a=2$ and $b=1$. Equation~(\ref{logFP})
defines what is known as the fundamental plane (FP) of elliptical
galaxies, a scaling between the three global properties of mass --
implicit in the stellar surface brightness --, size, and velocity
dispersion (or specific internal energy) that shape up the pivotal
space (hereafter the $RVM$ space) in which to study galaxies of all
kinds \citep*[see, e.g.][and references
  therein]{DS10,Tor11,BFvD15}\footnote{The exact definitions of the
  global variables used in these equations are irrelevant, as long as
  they refer to quantities that can be related directly to the
  corresponding properties entering the virial theorem.}. Introduced
by \citet{DD87} and \citet{Dre87} as a precise distance indicator for
ETGs, the widely studied FP of ellipticals combines the two well-known
2D \citeauthor*{FJ76} $L$--$\sigma$ relation \citetext{FJR;
  \citealt{FJ76}} and $R$--$\mu$ relation \citep{Kor77} into a 3D
scaling in $RVM$ space with a yet more reduced scatter ($< 0.1$ dex in
the perpendicular direction). Both the tilt and tightness of this
plane are used as powerful constraints on theoretical models of ETG
formation.

\citet{Rob06} also demonstrated that for sufficiently high gas
fractions ($f_{\rm gas} \ge 0.4$) the remnants of pairwise mergers can
yield a simulated FP with nearly the same tilt ($\lambda\sim 0.75$)
observed in the NIR \citetext{see, e.g., the recent results by
  \citealt{LaB10}; values quoted in the literature, however, may vary
  with wavelength, environment and redshift}. In the same way,
gas-rich dissipational simulations produce model $R$--$\ms$ relations
with a slope ($\mu\simeq 0.55$) that compares well with the relation
measured for nearby ETGs \citep[e.g.][]{She03,H-C13}, and that is
fully consistent with the formula~(\ref{lambda}) given the previously
quoted values of $\lambda$ and $\gamma$. Further analysis of these
sort of simulations by \citet{Cox06} and \citet{Hof10} showed that an
increase in the gas fraction produced remnants progressively more
oblate and rounder, that rotate faster and that, in general, match far
better than their dissipationless counterparts the 1D and 2D
structural and kinematic properties of $\sim 80\%$ of the nearby ETGs
observed in integral-field spectroscopic surveys \citep{Ems07}.
However, the same dissipational merger simulations that successfully
reproduced the observed properties of low- and intermediate-mass
ellipticals, had trouble explaining the formation of luminous,
non-rotating spheroids, with old stars, shallow cusps and boxy
isophotes, which make the remaining $20\%$ of the local ETG
population \citep[][]{Nov08,Moo14}. On the other hand, now a
growing body of evidence, both theoretical \citep[e.g.][]{She03} and
observational \citep[e.g.][]{Ber07,Liu08,Bei14}, indicates that the
mass growth of the largest ellipticals, especially the brightest
members of groups and clusters, is quite likely driven by a sequence
of (gradually minor) dissipationless mergers, the role of gas being
restricted at most to very early epochs
\citep[e.g.][]{DeLB07,Ose10}. More interesting still, is the fact that
first-ranked galaxies define essentially the same FP of their less
massive counterparts, although with an even smaller scatter
\citep[e.g.][]{Ber07,Liu08}. This reduced dispersion is also
in agreement with a dry merger formation history.

All these findings are consistent with the idea, advanced many years
before by \citet{Kor89} and \citet*{BBF92}, that the importance of gas
during the merger process declines with galaxy mass. This hypothesis,
which \citeauthor{BBF92} embodied in the concept of \emph{merging
  continuum}, seems none the less to be in conflict with the outcome
of most galaxy merger experiments. In many of these simulations the
tilt of the FP is exclusively driven by gas dissipation, with the role
of gravity being limited to the preservation of the slope created by
the gas \citep[e.g.][]{NLC03,Rob06}. If this were the case, then the
tilted FP of first-ranked galaxies in the local universe, which as
observations suggest are essentially composed of very old stars
(formed at $z>2$), would be a legacy from at least $\sim 10$ billion
years ago. We believe that the root of the problem is that, as
mentioned before, both dry and wet pre-prepared merger simulations
have traditionally focused on binary collisions, recreating merger
hierarchies at best throughout two or three non-overlapping merger
steps which use the remnants of the previous one to realize the
initial conditions of the next \citep{Dan03,Rob06,Nip09,Moo14}. It is
quite clear that these idealizations of multiple merging are only
gross descriptions of the diversity, in terms of frequencies and mass
ratios, of galaxy interactions that shape the brightest ETGs in the
course of cosmic evolution.

Pending the study of large sets of brightest members of galaxy systems
through full cosmological simulations becomes computationally
afordable, the best approach to investigate the assembly and
properties of such objects by means of cosmologically-consistent
merger histories is nowadays through controlled experiments of
aggregations of multiple galaxies undergoing gravitational
collapse. It is important to focus on groups of galaxies rather than
on pairs, because of the necessity of building a dense environment in
which galaxies are prone to suffer numerous collisions, whilst
dynamically-young small-size overdensities provide a framework in
which both the frequency and strength of interactions, and therefore
the merger likelihood, are maximized. To our knowledge, the only
investigation on the role of gravitational dynamics in the formation
of massive ETGs conducted along these lines is the recent work by
\citet*{TDY13,TDY15}. By studying collapsing groups of three to
twenty-five disc galaxies spanning a range of masses, these authors
have demonstrated that multiple dry merging is certainly a plausible
route for the formation of the most massive ellipticals. In
particular, \citeauthor{TDY15} have shown that hierarchical sequences
of dissipationless mergers of spiral galaxies can produce first-ranked
objects obeying a significantly tilted FP, although somewhat less
strongly than observations suggest. In the same way, they have found
that this scenario results in exceptionally tight 2D and 3D
relationships between the most important physical properties of
central remnants.

{\null In the next section, we present a similar experiment to that of
  \citeauthor{TDY13}, based on controlled high-resolution numerical
  simulations of the previrialization stage of isolated groups of
  galaxies, which produces even more striking outcomes regarding the
  origin of the FP defined by the most massive members of galaxy
  aggregations (Sec.~\ref{results}). Our set-up is somewhat more
  elaborated than theirs in certain aspects. The main differences in
  our approach -- in which the key of our better performance appears
  to lie -- are: the simultaneous inclusion of late- and early-type
  galaxy models to account for the expected mix of progenitors'
  populations; the non-homologous scaling of the progenitors' global
  properties; the inclusion of a common dark matter background within
  which galaxies are randomly distributed; and the scaling of the
  progenitors' dark halo masses to values appropriate for the initial
  redshift of the simulation, which is also significantly higher
  ($\zi=3$ vs $\zi=2$). By contrast, we are more restrictive than
  \citeauthor{TDY13} in terms of the range of both the initial
  membership and total masses of the simulated groups, which we fix to
  specific typical values (see below). In no case this latter
  constraint undermines the conclusions of the present work.}

\section{Simulations and methods}\label{simul}

{\null As part of a broader project aimed at studying galaxy merging in
  a realistic context, we have developed a high-resolution $N$-body
  model of a forming aggregation of galaxies which is a good and
  computationally competitive approximation to pure cosmological
  experiments.} Here we provide a summary of the main characteristics
of this tool and the methodology adopted to analyze its outcomes, both
discussed in great detail in \citeauthor{Sol16}
\citetext{\citeyear{Sol16}; hereafter \citeauthor*{Sol16}}.

\subsection{Group model}\label{grp_model}

Our simulated groups start at redshift $\zi=3$ as a (nearly) uniform,
bound spherical overdensity initially expanding. Assuming that
pressure gradients are negligible, a top-hat perturbation of amplitude
$\deli>0$ at the initial time $\ti$ evolves like a Friedmann universe
of initial radius given by
\begin{equation}
\Rp(\ti)=\left[\frac{3\Mp}{4\pi\rhoc(\ti)\Om(\ti)(1+\deli)}\right]^{1/3}\;,
\end{equation}
where the suffix 'p' identifies quantities associated with the
perturbation, i.e., the group, while the critical density,
$\rhoc(\ti)$, and the mass density parameter, $\Om(\ti)$, refer to the
unperturbed background cosmology -- the contribution of the
cosmological constant, $\Lambda$, to the evolution of the group can be
neglected. For the present work, we deal with simulated groups of
total mass $\Mp=M_{\rm{tot,gr}}=10^{13}\,h^{-1}\msun$ ($h\equiv
H_0/100$~km/s/Mpc), with the value of the initial overdensity,
$\deli$, chosen so that a perfectly homogeneous top-hat perturbation
of the same mass would collapse at $z=0$. In our experiments, the time
elapsed before turnaround, about $1/3$ of the total simulation time of
$\sim 11.5$ billion years in a standard concordant $\Lambda$CDM
cosmology, serves for gravitational interactions between group members
to build up peculiar velocities and $n$-point correlations among their
positions, thus guaranteeing a reasonable approximation to a
conventional galaxy group before merging becomes widespread during the
non-linear gravitational collapse phase.

\begin{figure*}
\begin{minipage}{175mm}
\centering
\includegraphics[angle=0,width=0.92\linewidth]{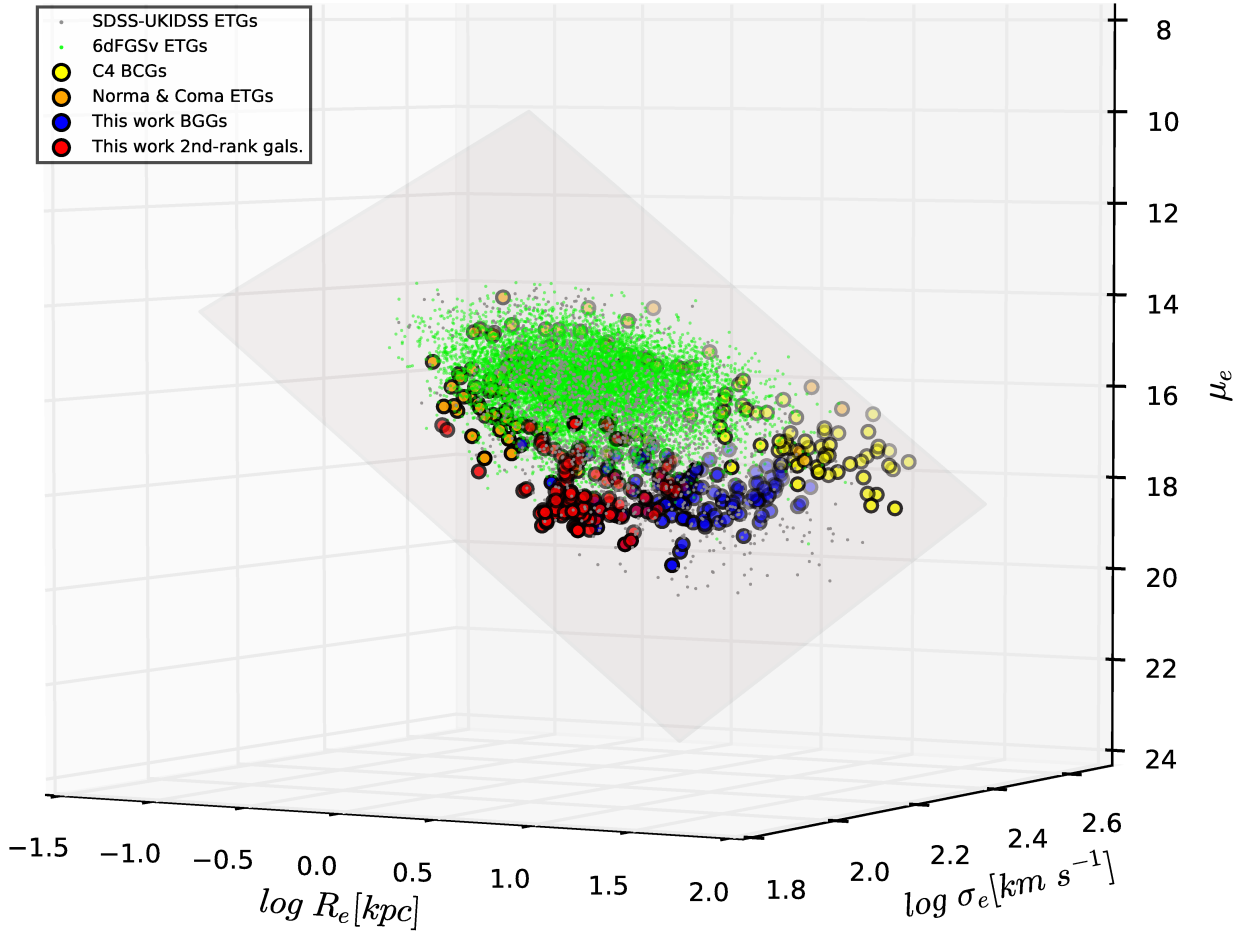}
\vspace{0pt}
\caption{\small The tilted FP of massive hot spheroids. 3D graphical
  representation of the FP (light-gray plane) defined by our
  mock first-ranked group galaxies (big dark-blue circles) in the
  space of parameters $[\log\Reff,\log\sige,\SBe ]$. The
  distribution of our less-reliably-resolved second-ranked galaxies is
  also shown (big red circles), together with four published data sets
  of nearby ETGs: a homogeneous set of 85 BCGs extracted from clusters
  (big yellow circles), the ETGs in the Norma and Coma clusters (big
  orange circles), the 6dFGSv catalog (green dots), and a
  volume-limited sample of ETGs based on SDSS-UKIDSS observations
  (small gray circles). Details on the transformations between
  simulated and observed measurements in this and the following
  figures can be found in \citeauthor*{Sol16}. All subsets
  included in the plot show a strikingly similar tilt and
  normalization. Note also that we do not draw error bars in our
  measurements, but plot instead three independent points for each
  remnant corresponding to estimates along the Cartesian axes.}
\label{figu_1}
\end{minipage}
\end{figure*} 

Each group region contains initially $\Ngal=25$ non--overlapping
extended galaxy haloes. The total (virial) masses of this individual
haloes, which obey a Navarro-Frenk-White density profile, are randomly
drawn from a Schechter probability distribution function (pdf) of
asymptotic slope $\alpha=-1.0$ and characteristic mass
$\mc=10^{12}\;h^{-1}\msun$. We set a lower limit in mass of
$0.05\;\mc$, as smaller haloes do not play a significant role in the
results. For a given halo mass, the remaining global parameters of the
dark haloes are fully determined by the background cosmology given a
time of observation ($\zo\simeq 0$). A $5\%$ of the total mass of each
halo is placed in a stellar (baryonic) core following either a
disc-bulge or a pure spheroidal distribution. Galaxies with masses
less than $0.1\mc$ are assumed for simplicity to host only spheroidal
stellar distributions. Above this mass threshold, galaxy morphologies
are established using a Monte Carlo technique that assumes a late-type
galaxy fraction of $0.7$. The initial structural and dynamical
properties of all central stellar distributions are set so that they
are consistent with the main local scaling laws defined by real
objects of the same class. For late-type galaxies, disc scale-lengths,
$\Rd$, are initialized according to the analytic formalism for disc
formation developed by \citet{DS10}, which relies on specific angular
momentum conservation during gas cooling and adiabatic halo response
to gas inflow. According to this model, which produces discs
consistent with the local Tully-Fisher relationship calculated by
\citet{Mas06}, $\Rd$ is equal to the virial radius of the parent halo
times a factor expressed in terms of elementary functions that, for a
given halo profile, depends on the values adopted for the halo
concentration, $c$, spin parameter, $\lambda$, and stellar mass
fraction, $f_\star$ (see \citealt{DS10} for a full account of the
scaling methodology). On the other hand, and given that the
adiabatic-gas-inflow model is not very suitable for classical
early-type galaxies, we set up such objects by directly relating their
size measure, represented by the effective radius of their projected
luminosity profiles, $\Reff$, to their total mass, $M$, via the
empirical formula
\begin{equation}\label{empirical_re}
\Reff = 2.05\;\left[\frac{M}{\mc}\right]^{5/8}\,h^{-1}\,{\rm{kpc}}\;.
\end{equation}
{\null This scaling matches the observed size-luminosity relation for
  Sloan Digital Sky Survey (SDSS) spheroids reported by \citet{Ber03}
  in the $i^*$-band if one assumes $f_\star=0.05$, a fixed rest-frame
  mass-to-light ratio $\Upsilon_I$ of $2.90\,h$ solar units, and a
  colour equation $(i^*-I)=0.53$ mag
  \citep*{FSI95}. Equation~(\ref{empirical_re}) is also consistent
  with the $\Reff$--$\ms$ relation inferred by \citet{She03}.}

After setting the central discy and spheroidal distributions of the
progenitor galaxies, we scale down the total masses and taper radius
of their dark haloes to values consistent with the initial redshift of
the simulations. The stellar cores, however, are left unchanged given
that in our experiments most mergers take place at a relatively late
epoch in which the properties of galaxies are rather similar to those
in the local volume. Once all the galactic haloes of a given group
have been generated and rescaled, the common uniform background of the
group is evenly filled with DM particles (identical to those that make
the individual haloes) until a total group mass (dark plus luminous)
of $10\;\mc$ is reached. The initial positions of background particles
(including the centre of mass of member galaxies) are randomly
distributed inside the group volume, while their initial velocities
follow the local Hubble flow.

\subsection{Analysis of simulation data}\label{analysis}

A total of 48 groups have been created following the procedure just
described. In order to compare mock and real data, we have projected
the 3D spatial information contained in the final snapshot of each run
on to the three Cartesian planes and, after adding a minimum sky
background to all images, smoothed them with an isotropic kernel. We
then have used the software SExtractor \citep{BA96} to build a catalog
of group members. The global properties of the brightest group
galaxies (BGGs) and of its lesser companions have been calculated from
the three independent projected final images resulting from each
simulation. This has allowed us to triple the sample size, offering at
the same time a crude but useful information on the variance in the
simulated data arising from object identification, projection effects
and contamination by foreground companions. Further details on the
description of the methodology adopted to process our simulations can
be found elsewhere (\citeauthor*{Sol16}).

For each group central remnant we have considered estimators of the
basic global properties of size, internal velocity dispersion and
scale, respectively represented by the half-stellar mass or effective
radius, $\Reff$, the mean stellar velocity dispersion within $\Reff$,
$\sige$, and in the latter case indistinctly by either $\ms$, the
total stellar mass, or by 
$\SBe \equiv -2.5\log[I_{\rm e}] = -2.5\log[(\ms/2)/(\pi\Reff^2)]$, 
the mean stellar mass surface density within $\Reff$. The use of the 
effective surface density in the definition of the FP makes this 
flat surface essentially independent of scale, because for spherically
symmetric virialized systems its three dimensions are proportional to
$M^{1/3}$. Another advantage of $\SBe$ is that it is a proxy for the
mean surface brightness from old stars -- measured in a near infrared
window of the electromagnetic spectrum, such as the $K$-band --, a
distance-independent property (for nearby objects) preferred by
observers.

\section{Results and discussion}\label{results}

\begin{figure}
\centering
\includegraphics[angle=0,height=0.86\textheight]{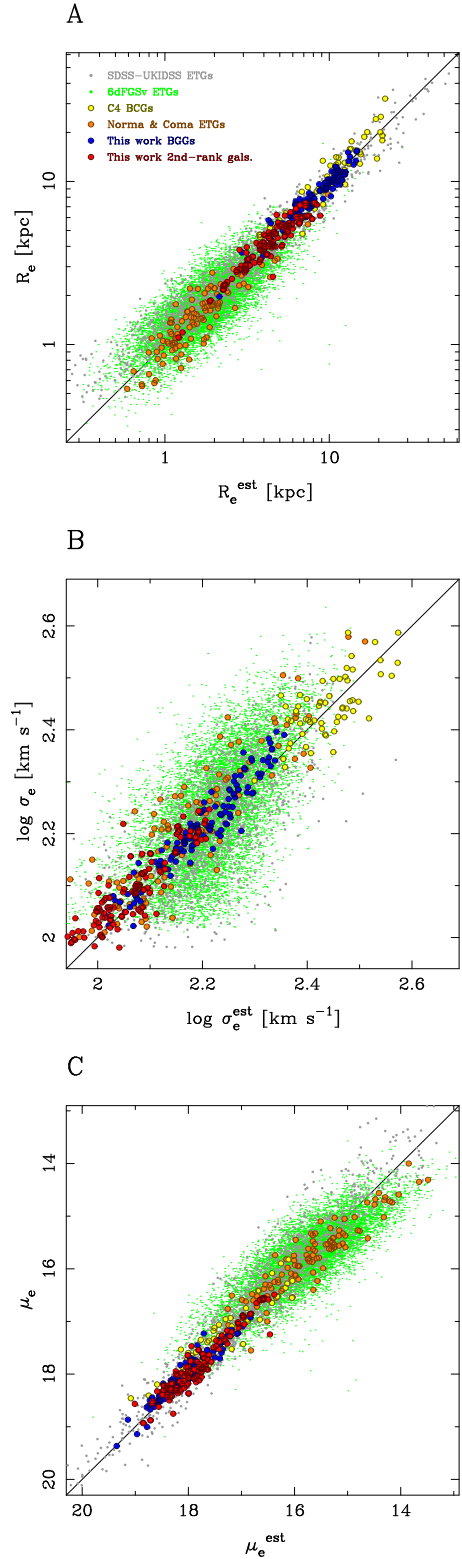}
\caption{\small Residuals of FP fitting along the dimensions of the
  standard plane. From top to bottom: (A) $\Reff$, (B) $\sige$, and
  (C) $\SBe$. Data symbols are the same used in Fig.~\ref{figu_1}. In
  all panels the black diagonal line represents perfect agreement
  between real ($Y$ axis) and estimated ($X$ axis) values. The
  reasonably neutral behaviour of the residuals confirms that the
  model FP fits the observations well. Note that the samples
  containing the brightest objects, both simulated and observed,
  systematically show the smallest differences between expectations
  and measurements.}
\label{figu_2}
\end{figure}

In Fig.~\ref{figu_1} we provide a 3D graphical representation of the
orthogonal (stellar mass) FP defined by our mock first-ranked group
galaxies in the global parameter space $[\log\Reff,\log\sige,\SBe]$. 
The equation of the best-fitting orthogonal regression model is
\begin{equation}
\begin{split}
\log\Reff = & (1.60\pm 0.05)\log\sige\\ 
 & + (0.75\pm 0.02)\SBe/2.5 -(8.0\pm 0.2)\;,
\end{split}
\end{equation}
which is tilted relative to the standard plane defined by virialized
systems that are homologous in a broad sense (see the Introduction).
Quoted confidence intervals are median absolute deviations from
$4,000$ bootstraps. In this figure we also plot the homogenized
$K$-band data \citetext{see \citeauthor*{Sol16}} from four different
public compilations of massive hot spheroidal systems: a homogeneous
set of 85 BCGs, the cluster counterparts of BGGs, with redshift $z$
less than $0.1$ extracted from the C4 cluster catalog \citep{Liu08},
the recently compiled catalog of ETGs in clusters Norma and Coma by
\citet{Mut14}, the 6dFGS Fundamental Plane (6dFGSv) catalog of nearly
$9,000$ ETGs in the local ($z < 0.055$) universe \citep{Cam14}, and a
sample of $1,430$ ETGs in the redshift range $0.05$--$0.095$ which
combines photometric data from the SDSS and the UKIRT Infrared Deep
Sky Survey (UKIDSS) \citep{LaB08}. The FP delineated by our simulated
remnants matches perfectly well not just the mean tilt, but also the
normalization of the relatively narrow, flat 3-space region defined by
the observations, all without requiring that progenitor galaxies have
gas. This can be better visualized in Fig.~\ref{figu_2} which compares
the predicted values for each variable defining the inferred FP with
the actual measurements. The lack of a clear non-random structure in
the distribution of residuals is a sign that the model FP fits all the
observations well. A collateral result of this analysis is the
identification of a set of completely independent databases of giant
ellipticals with fully compatible measurements despite the not too
insignificant differences which affect the selection and treatment of
the data. This comes on top of recent claims that the FP of massive
galaxies does not vary with environment, redshift or even
star-formation activity \citep{LaB08,Mag12,BFvD15}. The successful
reproduction of the tilted FP relation is in sharp contrast with
former dissipationless merger experiments that, with a few exceptions
\citep{AV05,TDY15}, have systematically produced remnants which occupy
a FP similar to that expected from virialized homologous systems. To
our knowledge, ours is the most accurate experimental prediction of
this basic relationship, regarding both its three coefficients and its
reduced thickness, ever inferred from a numerical simulation of any
kind aimed at studying the formation of the most massive galaxies in
the universe.

Another notable feature of the FP is that different classes of giant
ETGs occupy distinctive areas on its surface, reflecting diversity in
the structural and kinematic properties related to their respective
formation histories \citep{Fab97,Ber07,BFvD15}. First-ranked galaxies
are clearly segregated from the locus of the general population of
giant ellipticals because of their considerably lower surface stellar
mass density/brightness \citep{MenA12}, whereas second-ranked objects
lie on an intermediate position -- being mindful that in simulations
the recovering of galaxy properties degrades with decreasing
brightness, we have found that mock galaxies with lower ranks depart
progressively from the FP. Calculation of the vectors defined by the
most important physical processes ruling the formation of hot stellar
systems (energy dissipation, dissipationless mergers, mass loss
through either supernova-driven winds or tidal and ram pressure
stripping, etc) in the full 3-space of the FP, have shown that they
move galaxies within the plane far more than they move them out of it
\citep{BBF92}. This provides a plausible explanation for the
remarkable fact that in said space the common manifold of giant ETGs
forms quite a slender sheet despite the strong stochasticity of the
galaxy formation process. Fig.~\ref{figu_2} also shows that the scatter
around the FP is similarly connected to the ETG class, becoming
smaller for classes with more massive representatives (this statement
is reflected in quantitative terms in Table 3 of 
\citeauthor*{Sol16}). The reduced scatter of \citeauthor{Liu08}'s BCG
data set, surpassed only by the even lower dispersion shown by our
simulated galaxies, is another manifestation of the important
regularity in the formation process of the most massive hot spheroids
in the universe which, according to our experiments, can be driven by
purely gravitational phenomena. 

The results outlined here, in harmony with other investigations of
stochastic collisionless merging \citep{TDY15}, demonstrate that this
mechanism is perfectly capable of reproducing the tight local
relations that show the most fundamental parameters of these
systems. It appears then that a realistic merger hierarchy can
naturally provide the conditions required to reproduce the small
intrinsic scatter shown by the observed stellar-mass scaling laws of
massive ETGs without the need for a high degree of fine tuning between
progenitors' mass ratios and merger orbital parameters
\citep{Nip09}. In particular it should be noted that our progenitor
galaxies have virial masses randomly drawn from a wide range of a
Schechter pdf and initial positions that are uniformly distributed
inside the group volume. Furthermore, the fact that we simulate groups
that are in the making implies that their member galaxies move most of
the time in a mean gravitational field that is widely fluctuating and
erratic because of the violent relaxation of the parent system. As a
result, the outcome of the mergers that take place in such a highly
non-linear phase -- more specifically, the statistical distribution of
the properties of the remnants -- gets largely detached from the
initial orbital characteristics of progenitor galaxies.

\begin{figure*}
\begin{minipage}{175mm}
\centering
\includegraphics[angle=0,width=\linewidth]{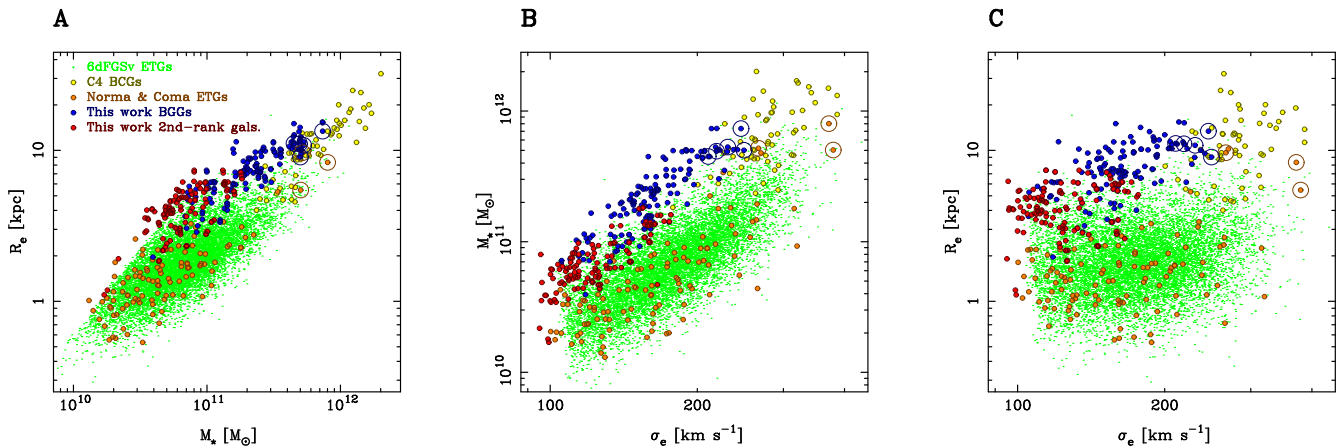}
\caption{\small Projections of the FP in $RVM$ space. (A) Size-stellar
  mass relation. (B) Stellar mass-velocity dispersion or mass FJR. (C)
  Size-velocity dispersion relation. Data symbols are the same used in
  Fig.~\ref{figu_1} except for the SDSS-UKIDSS ETGs \citep{LaB08},
  which have not been included because they lack stellar mass
  information. The data points corresponding to our five simulated
  BGGs showing the largest magnitude gap between them and the second
  brightest group companion, as well as the three CGs of the Norma (1
  object) and Coma (2 objs.) clusters, are encircled. The positions of
  these very big objects in the FP projections are fully compatible
  with the point cloud of BCGs (big yellow circles).}

\end{minipage}
\label{figu_3}
\end{figure*} 

The fact that all massive ETGs fall into a single FP, coupled with the
segregation of the highest-ranked objects from lesser galaxies,
entails that any non-edge-on projection of this flat surface must also
lead to segregated 2D scaling laws. 
To investigate this issue we now
use $\ms$ to represent scale, shifting to the (logarithmic) $RVM$
coordinate system defined by the three most basic global parameters
connected by the standard plane. In this manner we set the framework
for the study of two of the most firmly established empirical scaling
relations of elliptical galaxies: the \citeauthor*{Kor77}-like $RM$
relation \citep{She03} and the $VM$ relation, a mass analogue of the
classical FJR. In good agreement with observational studies of
BCG/BGGs \citep{Lau07,Ber07,Ber09,MenA12}, we find that our simulated
first-ranked galaxies lie off the standard 2D scaling relations
defined by the bulk of the ETG population. In particular, as shown in
Fig.~\ref{figu_3}, BGGs are larger and have lower effective velocity
dispersions than ordinary ellipticals of the same stellar mass. The
combination of this latter result with the fact that light appears to
be similarly concentrated for BGGs than for non-BGGs
(\citeauthor*{Sol16}), tells us that the total mass-to-light fraction
interior to $\Reff$ is lower for the former than for the latter. Our
numerical experiments, however, do not seem to support claims of the
steepening of the $R\propto M^{\alpha}$ relation for CGs toward values
$\alpha\gtrsim 1$ \citep{Lau07,Ber09}. We find instead that our
first-ranked objects, which occupy a locus moderately offset from the
central axis of the observational data in Fig.~\ref{figu_3}A, are well
fitted by a model with $\alpha=0.60\pm 0.03$. This value of the
power-law index agrees well with the slope $\alpha\sim 0.65$ found for
non-BCG galaxies in the Sloan $r$-band \citep{She03,Ber07}, while a
plain orthogonal fit to the general elliptical population data
included in Fig.~\ref{figu_3}A (green dots and small gray circles) also
gives $\alpha\sim 0.6$. Approaching closer to our findings,
\citet{Liu08} obtain $\alpha\sim 0.9$ from surface measurements up an
isophotal limit of 25 $r$-mag arcsec$^{-2}$, which reduces to $\sim
0.8$ when their magnitudes are transformed into mass in old stars,
whereas they get $\alpha\sim 0.75$ for a control sample of
non-BCGs. Differences in the photometry, the waveband of the
observations, sample construction (i.e.\ incompletenesses and
selection biases), and fitting methods would help explain the lack of
a closer agreement between these results about the most robust of the
2D relationships.

\begin{figure*}
\begin{minipage}{175mm}
\centering
\includegraphics[angle=0,width=\linewidth]{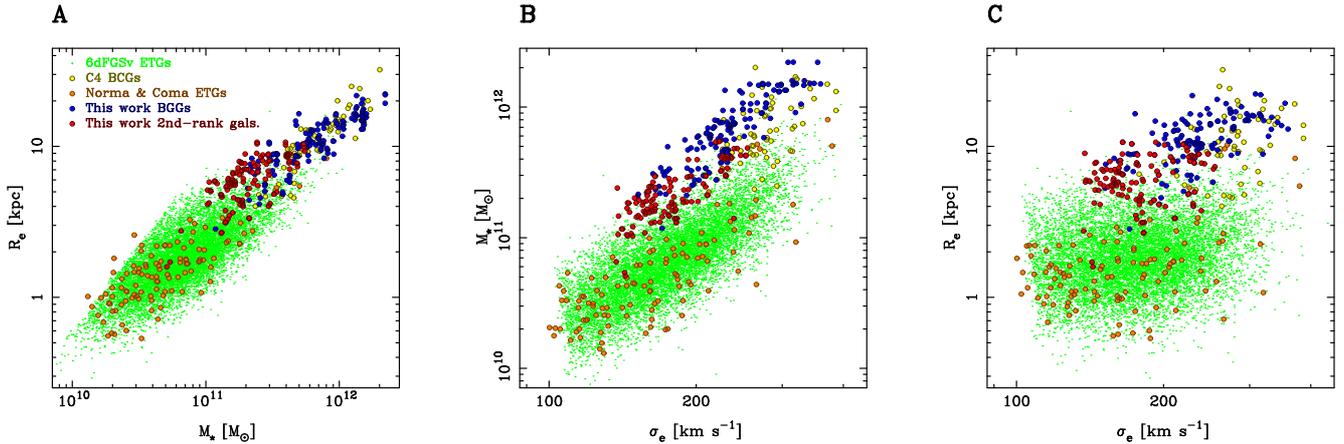}
\caption{\small Predicted joint distributions of the properties of the
  brightest galaxies in more massive groups. Same as in
  Fig.~\ref{figu_3} but after rescaling the properties of our simulated
  galaxies to the values expected when the mass unit of our
  simulations triples. The point clouds corresponding to our mock BGGs
  (big dark-blue circles) and the observed BCGs (big yellow circles)
  show now a substantial overlap in all three panels.}
\label{figu_4}
\end{minipage}
\end{figure*} 

The different locations of massive and regular ETGs along the FP also
have a bearing on the zero point of the $M\propto\sigma^\beta$
relation leaving, as in the previous case, the power law index hardly
affected. The joint $VM$ distribution (Fig.~\ref{figu_3}B) evidences
that our simulated first-ranked galaxies have excess stellar mass
above the prediction of the standard FJR, defining a roughly parallel
sequence $\sim 1.6$ mag brighter (more massive) whose upper boundary
extends into the area occupied by the BCGs' data. This large intercept
offset implies that, while the $RM$ relation is basically an edge-on
projection of the mass FP, the baryonic FJR is definitely
not. Besides, it may also help to explain why the values of the slope
of the FJR recovered from data sets that include a mix of BCGs and
non-BCGs are considerably steeper than the canonical $\beta=4$
\citep[e.g.][]{Lau07}. In the current case, we find $\beta = 3.74\pm
0.14$ from an orthogonal fit to the subset of our mock first-ranked
galaxies. This value is consistent with most measurements reported in
the literature \citep*{KD89,PDdC98,Koc06}.

Fig.~\ref{figu_3}C displays the joint distribution of $\Reff$ and
$\sige$, which shows the FP almost face-on, providing a clear view of
the almost non-existent correlation shown by these two variables --
illustrating the general intransitivity of correlations --, as well as
of the substantially different areas occupied by central and
non-central ETGs. This plot also best illustrates the moderate overlap
shown by the point clouds corresponding to our simulated BGGs and
their observational counterparts. This discrepancy can be ascribed to
the fact that our mock first-ranked galaxies probably represent the
lowest-mass objects of its class -- not all of them, for instance, are
truly massive: $\ms>10^{11}\msun$ -- since these are central remnants
assembled in isolated small groups rather than in the considerably
larger galaxy aggregations hosting the bulk of the observed
CGs. Indeed, as shown in the different panels of Fig.~\ref{figu_3}, the
data points encircled in black corresponding to five of our largest
simulated BGGs (the most fossil-like objects), as well as to the three
CGs of the Norma and Coma clusters, fall squarely into the observed
BCGs' region (big yellow dots) -- although on opposing sides --,
indicating that they all share similar characteristics. Pending that
we extend the present analysis to more massive galaxy aggregations, a
good way to substantiate the proposed explanation is by taking
advantage of the fact that our simulations are nominally scale-free,
as long as our preferred transformations to physical units adopted in
the initial setup continue to be acceptable. Thus, we have used the
standard $R\sim V\sim M^{1/3}$ scaling verified by virialized
spherical systems to infer a rough estimate of the 2D correlations
that are to be expected from our simulations when, for instance, the
mass of the host units triples (keeping $N_{\rm gal}$ constant). As
depicted by Fig.~\ref{figu_4}, our predictions indicate that the global
properties of simulated first-ranked galaxies of larger aggregations
would match much better the volume occupied by \citeauthor{Liu08}'s
BCGs in $RVM$ space, showing a substantial overlap along all three
projections. Interestingly, this would be accomplished virtually
without changing the slopes of the relations delineated by the
original data in Fig.~\ref{figu_3} (as mentioned earlier, in
$[\Reff,\sige,\SBe ]$-space any re-scaling of the mass has a null
impact on the FP coefficients).

Of all CG data sets used to date to investigate the scaling relations
of these objects, that by \citet*{Vul14}, based on an X-ray-selected
sample of 169 groups at intermediate redshift ($z\sim 0.6$) and virial
masses in the range $\sim[10^{13}$--$10^{14}]\,\msun$, is among the
most similiar to ours in terms of the mass of the parent
systems. Focusing again on comparisons mainly qualitative in nature,
we find that the joint distributions of size and stellar mass of both
BGG samples show a very good statistical agreement, in the sense that
\citeauthor{Vul14}'s straight-line fits to their data also provide
quite a reasonably good description of our point cloud
(Fig.~\ref{figu_3}A). In contrast, the other two relations involving
the central velocity dispersion of galaxies show large offsets. The
deviation is due to the fact that the intermediate-$z$ BGGs have
systematically higher values of $\sige$ (at fixed $\ms$ or $\Reff$)
than their synthetic counterparts. It is, however, important to
realise that the CGs in \citet{Vul14} with a measured velocity
dispersion constitute a magnitude-limited subset of the main
sample. This biases their measurements against faint, low-stellar-mass
objects, suggesting that the rescaled data from our Figs.~\ref{figu_4}B
and C would perhaps allow a more proper comparison. When doing so, we
find indeed a substantial improvement in the overall consistency of
the dynamic ranges in size, velocity and mass from both
samples. \citeauthor{Vul14}'s study also brings out the fact that the
late-time evolution of first-ranked galaxies eventually leads to
changes in their global properties that largely involve a
renormalization of the main 2D scaling laws that does not alter -- or
scarcely changes at worst -- their slopes (compare their Figs.\ 10 and
11). This behaviour, analogous to that of the mass rescaling applied
to our BGGs, suggests the stability of the FP with respect to the
secular evolution of CGs' properties, in good agreement with the
theoretical predictions about the nature of this surface by
\citet{BBF92} and the observations by \citet{Bez13}.

\section{Conclusion}\label{conclusion}

Controlled hydrodynamical simulations of galaxy merging indicate that
dissipation appears to be both necessary and sufficient to explain the
tilt of the FP of ETGs \citep[e.g.][]{Rob06,HCH08}. This puts under
strain observations of massive early-type objects claiming otherwise
\citep*{She03,NKB06,Ber07,Liu08} and therefore the conjecture,
advanced for more than a few years now by \citet{Kor89} and
\citet{BBF92}, that ellipticals as a group constitute a continuum of
merger-dominated systems in which the importance of gas in the outcome
declines with increasing galaxy mass. It has to be noted, however,
that the merger hierarchy typical of $\Lambda$CDM cosmologies has been
traditionally recreated in numerical experiments through a sequence of
two or three non-overlapping major binary collisions of homologous
systems that use the remnants of one step to realize the initial
conditions of the next \citep[e.g.][]{Dan03,Rob06,Nip09,Moo14}. These
idealizations of multiple merging can hardly capture the diversity of
galaxy interactions that take place in the course of cosmic evolution,
especially when it comes to the distribution of orbits, frequencies
and mass ratios. In contrast, our cosmologically-consistent
simulations of previrialized galaxy associations provide the most
convincing evidence presented to date on the viability of hierarchical
collisionless merging as a formation channel for first-ranked objects
through the reproduction of their main global properties and
correlations, leading therefore to the removal of the last remaining
obstacle for the experimental validation of the merger continuum
conjecture.

\section*{Acknowledgements}

We thank the referee, Carlo Nipoti, for his useful comments and
suggestions. This work was supported by the Program for Promotion of
High-Level Scientific and Technical Research of Spain under projects
AYA2010-15169 and AYA2013-40609-P.





\begin{thebibliography}{99}

\bibitem[Aceves \& Vel\'azquez(2005)]{AV05} Aceves H., Vel\'azquez H., 2005, \mnras, 360, L50
\bibitem[Beifiori et al.(2014)]{Bei14} Beifiori A.\ et al., 2014, \apj, 789, 92 
\bibitem[Bender et al.(1992)Bender, Burstein \& Faber]{BBF92} Bender R., Burstein D., Faber S.~M., 1992, \apj, 399, 462
\bibitem[Bernardi(2009)]{Ber09} Bernardi M., 2009, \mnras, 395, 1491
\bibitem[Bernardi et al.(2003)]{Ber03} Bernardi M.\ et al., 2003, \aj, 125, 1866
\bibitem[Bernardi et al.(2007)]{Ber07} Bernardi M., Hyde J.~B., Sheth R.~K., Miller C.~J., Nichol R.~C., 2007, \aj, 133, 1741
\bibitem[Bertin \& Arnouts(1996)]{BA96} Bertin E., Arnouts S., 1996, \aap, 317, 393
\bibitem[Bezanson et al.(2013)]{Bez13} Bezanson R., van Dokkum P.~G., van de Sande J., Franx  M., Leja J., Kriek M., 2013, \apj, 779, L21
\bibitem[Bezanson et al.(2015)Bezanson, Franx \& van Dokkum]{BFvD15} Bezanson R., Franx M., van Dokkum P.~G., 2015, \apj, 799, 148
\bibitem[Campbell et al.(2014)]{Cam14} Campbell L.~A.\ et al., 2014, \mnras, 443, 1231
\bibitem[Cappellari et al.(2013)]{Cap13} Cappellari M.\ et al., 2013, \mnras, 432, 1862
\bibitem[Carlberg(1986)]{Car86} Carlberg R.~G., 1986, \apj, 310, 593
\bibitem[Ciotti(1997)]{Cio97} Ciotti L., 1997, in da Costa L.N., Renzini, A., eds., Galaxy Scaling Relations: Origins, Evolution and Applications. Springer-Verlag, Berlin, p.\ 38
\bibitem[Ciotti et al.(2007)Ciotti, Lanzoni \& Volonteri]{CLV07} Ciotti L., Lanzoni B., Volonteri M., 2007, \apj, 658, 65
\bibitem[Cox et al.(2006)]{Cox06} Cox T.~J., Dutta S.~N., Di Matteo T., Hernquist L., Hopkins P.~F., Robertson B., Springel V., 2006, \apj, 650, 791
\bibitem[Dantas et al.(2003)]{Dan03} Dantas C.~C., Capelato H.~V., Ribeiro A.~L.~B., de Carvalho R.~R., 2003, \mnras, 340, 398
\bibitem[Darriba \& Solanes(2010)]{DS10} Darriba L., Solanes J.~M., 2010, \aap, 516, A7
\bibitem[De Lucia \& Blaizot(2007)]{DeLB07} De Lucia G., Blaizot J., 2007, \mnras, 375, 2
\bibitem[Djorgovski \& Davis(1987)]{DD87} Djorgovski S., Davis M., 1987, \apj, 313, 59 
\bibitem[Dressler et al.(1987)]{Dre87} Dressler A., Lynden-Bell D., Burstein D., Davies R.~L., Faber S.~M., Terlevich R., Wegner G., 1987, \apj, 313, 42
\bibitem[Eggen et al.(1962)Eggen, Lynden-Bell \& Sandage]{EL-BS62} Eggen O.~J., Lynden-Bell D., Sandage A.~R., 1962, \apj, 136, 748
\bibitem[Emsellem et al.(2007)]{Ems07} 
Emsellem E.\ et al., 2007, \mnras, 379, 401
\bibitem[Faber \& Jackson(1976)Faber--Jackson]{FJ76} Faber S.~M., Jackson R.~E., 1976, \apj, 204, 668
\bibitem[Faber et al.(1997)]{Fab97} Faber S.~M.\ et al., 1997, \aj, 114, 1771
\bibitem[Fukugita et al.(1995)Fukugita, Shimasaku \& Ichikawa]{FSI95} Fukugita M., Shimasaku K., Ichikawa T., 1995, \pasp, 107, 945
\bibitem[Hoffman et al.(2010)]{Hof10} Hoffman L., Cox T.~J., Dutta S., Hernquist L., 2010, \apj, 723, 818
\bibitem[Hopkins et al.(2008)Hopkins, Cox \& Hernquist]{HCH08} Hopkins P.~F., Cox T.~J., Hernquist L., 2008, \apj, 689, 17
\bibitem[Huertas-Company et al.(2013)]{H-C13} Huertas-Company M.\ et al., 2013, \mnras, 428, 1715
\bibitem[Kochanek(2006)]{Koc06} Kochanek C.~S., 2006, in Meylan G., Jetzer P., North P., eds., Saas-Fee Advanced Course Num.\ 33. Springer-Verlag, Berlin, p.\ 91
\bibitem[Kormendy(1977)]{Kor77} Kormendy J., 1977, \apj, 218, 333 
\bibitem[Kormendy(1989)]{Kor89} Kormendy J., 1989, \apj, 342, L63 
\bibitem[Kormendy \& Djorgovski(1989)]{KD89} Kormendy J., Djorgovski S., 1989, \araa, 27, 235 
\bibitem[La Barbera et al.(2008)]{LaB08} La Barbera F., Busarello G., Merluzzi P., de la Rosa I.~G., Coppola G., Haines C.~P., 2008, \apj, 689, 913
\bibitem[La Barbera et al.(2010)]{LaB10} La Barbera F., de Carvalho R.~R., de La Rosa I.~G., Lopes P.~A.~A., 2010, \mnras, 408, 1335
\bibitem[Lauer et al.(2007)]{Lau07} Lauer T.~R.\ et al., 2007, \apj, 664, 226
\bibitem[Liu et al.(2008)]{Liu08} Liu F.~S., Xia X.~Y., Mao S., Wu H., Deng Z.~G., 2008, \mnras, 385, 23
\bibitem[Magoulas et al.(2012)]{Mag12} Magoulas C.\ et al., 2012, \mnras, 427, 245 
\bibitem[Masters et al.(2006)]{Mas06} Masters, K.~L., Springob, C.~M., Haynes, M.~P., Giovanelli, R., 2006, \apj, 653, 861
\bibitem[M\'endez-Abreu et al.(2012)]{MenA12} M\'endez-Abreu J.\ et al., 2012, \aap, 537, A25
\bibitem[Moody et al.(2014)]{Moo14} Moody C.~E., Romanowsky A.~J., Cox T.~J., Novak G.~S., Primack J.~R., 2014, \mnras, 444, 1475
\bibitem[Mutabazi et al.(2014)]{Mut14} Mutabazi T., Blyth S.~L., Woudt P.~A., Lucey J.~R., Jarrett T.~H., Bilicki M., Schr\"oder A.~C., Moore S.~A.~W., 2014, \mnras, 439, 3666
\bibitem[Naab et al.(2006)Naab, Khochfar \& Burkert]{NKB06} Naab T., Khochfar S., Burkert A., 2006, \apj, 636, L81 
\bibitem[Nipoti et al.(2003)Nipoti, Londrillo \& Ciotti]{NLC03} Nipoti C., Londrillo P., Ciotti L., 2003, \mnras, 342, 501
\bibitem[Nipoti et al.(2009)]{Nip09} Nipoti C., Treu T., Auger M.~W., Bolton A.~S., 2009, \apj, 706, L86
\bibitem[Novak(2008)]{Nov08} Novak, G.~S.\ 2008, PhD thesis, Univ.\ of California, Santa Cruz 
\bibitem[Oser et al.(2010)]{Ose10} Oser L., Ostriker J.~P., Naab T., Johansson P.~H., Burkert A., 2010, \apj, 725, 2312
\bibitem[Ostriker(1980)]{Ost80} Ostriker J.~P., 1980, Comments on Astrophysics, 8, 177
\bibitem[Padmanabhan et al.(2004)]{Pad04} Padmanabhan N.\ et al., 2004, New Astronomy, 9, 329
\bibitem[Pahre et al.(1998)Pahre, Djorgovski \& de Carvalho]{PDdC98} Pahre M.~A., Djorgovski S.~G., de Carvalho R.~R., 1998, \aj, 116, 1591
\bibitem[Robertson et al.(2006)]{Rob06} Robertson B., Cox T.~J., Hernquist L., Franx M., Hopkins P.~F, Martini P., Springel V., 2006, \apj, 641, 21
\bibitem[Shen et al.(2003)]{She03} Shen S., Mo H.~J., White S.~D.~M., Blanton M.~R., Kauffmann G., Voges W., Brinkmann J., Csabai I., 2003, \mnras, 343, 978
\bibitem[Solanes et al.(2016)Paper~I]{Sol16} Solanes J.~M., Perea J.~D., Darriba L., Garc\'\i a-G\'omez C., Bosma A., Athanassoula E., 2015, \mnras, submitted (Paper~I)
\bibitem[Taranu et al.(2013)Taranu, Dubinski \& Yee]{TDY13} Taranu D.~S., Dubinski J., Yee H.~K.~C., 2013, \apj, 778, 61
\bibitem[Taranu et al.(2015)Taranu, Dubinski \& Yee]{TDY15} Taranu D.~S., Dubinski J., Yee H.~K.~C., 2015, \apj, 803, 78
\bibitem[Toomre(1977)]{Too77} Toomre A., 1977, in Tinsley B.~M., Larson R.~B., eds., Evolution of Galaxies and Stellar Populations. Yale University Observatory, New Haven, p.\ 401
\bibitem[Toribio et al.(2011)]{Tor11} Toribio M.~C., Solanes J.~M., Giovanelli R., Haynes M.~P., Martin A.~M., 2011, \apj, 732, 93
\bibitem[Trujillo et al.(2011)Trujillo, Ferreras \& de La Rosa]{TFdR11} Trujillo I., Ferreras I., de La Rosa I.~G., 2011, \mnras, 415, 3903
\bibitem[Vulcani et al.(2014)]{Vul14} Vulcani B.\ et al, 2014, \apj, 797, 62
\bibitem[Weil \& Hernquist(1996)]{WH96} Weil M.~L., Hernquist L., 1996, \apj, 460, 101

\end{thebibliography}








\bsp	
\label{lastpage}
\end{document}